Rapid $^{14}$C excursion at 3372-3371 BCE not observed at two different locations – comment on Wang *et al.* (2017)

A J Timothy Jull[1,2], Irina P. Panyushkina[3], Mihály Molnár[2], Tamás Varga[2], Lukas Wacker, Nicolas Brehm[4], Chris Baisan[3], Matthew W. Salzer[3] and Willy Tegel[5].

1. Department of Geosciences, University of Arizona, Tucson, Arizona USA
2. Isotope Climatology and Environmental Research Centre, Institute for Nuclear Research, Debrecen, Hungary.
3. Laboratory for Tree-Ring Research, University of Arizona, Tucson, Arizona USA.
4. Laboratory of Ion Beam Physics, ETH-Zürich, Switzerland.
5. Chair of Forest Growth and Dendroecology, University of Freiburg, Germany.

**Excursions in the carbon-14 record measured in tree rings are attributed to various high energy but short-lived cosmic effects [1-7]. So far, rapid changes at 774-775 CE, 993-994 CE and 660 BCE have been convincingly interpreted as due to rapid changes in solar cosmic-ray flux, usually accredited with reproduction of the events in trees at different locations and different laboratories, and followed with equivalent peaks of other cosmogenic isotopes in polar ice cores [8,9]. In 2017, Wang *et al.* [10] proposed a new event at 3372-3371 BCE based on a single set of annual $^{14}$C data measured on a floating series of tree rings sampled from a buried specimen of Chinese wingnut tree (*Pterocarya stenoptera*). We attempted to reproduce this event in tree rings of an absolutely dated bristlecone pine specimen (*Pinus longaeva*) from the White Mountains, California USA and a subfossil European oak (*Quercus sp.*) from the Moselle River Valley in France. Unfortunately, we are unable to determine whether there is any rapid $^{14}$C excursion during this time, although the regular 11-year Schwabe cycle is clearly observable. Therefore, the presence of a cosmic-ray event as suggested at 3372-3371 BCE cannot be confirmed. It is possible the exposed $^{14}$C spike occurred at some other time, the absolute age of which we cannot determine unless the precise calendar age of the Chinese wingnut wood is properly crossdated, although the other data of Wang *et al.* [10] is not inconsistent with our records. In either case, this is important because the frequency of such extreme cosmic events needs to be well-documented to assess the likelihood of similar high-energy effects occurring in the future.**

The well-known radionuclide carbon-14 ($t_{1/2}$ 5,730 years) is produced by the reaction of secondary thermal neutrons derived from the primary cosmic-ray flux on nitrogen [11]. This $^{14}$C is incorporated into the carbon cycle within 1-2 years as is most convincingly demonstrated through the development of the $^{14}$C signal from anthropogenic nuclear testing [12]. Miyake *et al.* [2, 3,4] were the first to demonstrate that studying the annual signal of $^{14}$C in tree rings revealed annual excursions which were much larger than those observable in the decadally-averaged international radiocarbon calibration curve (IntCal13 [13]). Specifically, they observed spikes in $^{14}$C activity at 774-775 CE and 993-994 CE [2, 4]. The 774 CE event was first confirmed in a bristlecone pine record [5] and indeed, this work encouraged many subsequent studies looking for both these events and searches for other events. Büntgen *et al.* [1] have summarized an extensive record from studies of many different trees for 774-775 CE from 34 locations and another set of trees for 993-994 CE from 10 locations around the globe. Another rapid event at 660 BCE has been reproduced in different records and is therefore also widely accepted [7, 9]. Separately, other rapid changes which may show more complex solar dynamo phenomena, or combinations of solar and galactic events at 5480 BCE and 813 BCE have also been observed [14,15].

An important task in demonstrating a convincing global signature of a $^{14}$C excursion is to reproduce the event with tree rings from different geographic locations. In this paper, we attempted to reproduce the $^{14}$C sequence of Wang et al. [10] using two independent tree-ring records. First, we measured $^{14}$C on tree-rings of bristlecone pine *(Pinus longaeva)* from the White Mountains in California (USA). This tree species is well-known for a remarkably long life span often exceeding several thousand years. The SH146-2003 remnant specimen was collected at the high-elevation Sheep Mountain site and rigorously crossdated with the Sheep Mountain master tree-ring chronology which spans 4408 BCE to 2014 CE [16]. In addition, we measured another independently developed $^{14}$C series from tree rings of subfossil European oak (*Quercus sp.*) collected from a gravel pit near Champey-sur-Moselle (France). The CHEY1-17 oak specimen is absolutely cross-dated with the South German oak chronology covering 8240 BCE – 2017 CE [17].

**Results**

We obtained $^{14}$C measurements on tree-ring samples from 1) bristlecone pine for the interval 3351-3392 BCE (42 years) and 2) European oak covering the period 3350-3390 BCE (40 years). The bristlecone pine specimen used for the $^{14}$C measurements has 328 years (3598-3271 BCE) and the European oak record has 93 tree rings (3402-3310 BCE). The tree rings are absolutely dated via crossdating with original site (master) chronologies. The Pearson correlation between the tree-ring width series of SH146-2003 and CHEY1-17 and their master tree-ring chronologies is 0.57 (p<0.01) and 0.52 (p<0.01), respectively. Individual ring samples were converted to α-cellulose and then combusted to $CO_2$ and reduced to graphite for analysis by accelerator mass spectrometry (AMS) as discussed in the methods. This period was chosen so that we could observe if the Wang *et al.* [10] event occurred during the same sampling period of 3358-3388 BCE expected from those results.

Our results are shown in figure 1 and supplementary table 1 that plot $\Delta^{14}$C [18] against known age of the tree rings. Results show an apparent periodicity of ~11 years with an amplitude of ~5 per mil on a declining trend of $\Delta^{14}$C. We note that the two $^{14}$C series, from bristlecone pine and European oak, are well-correlated with a Pearson correlation coefficient of 0.894. A student's t-test of these two data sets give good agreement with t=-0.686 (p=0.49, Tc=1.99). We also compared our results to those of Wang *et al.* [10]. Intriguingly, the Chinese wingnut $^{14}$C measurements agree with the general trend of our results, except for the two years of 3370-3371 BCE, where Wang *et al* [10] reported the new $^{14}$C excursion, and the year 3388 BCE. However, the excursion proposed by Wang *et al.* [10] cannot be confirmed in the other two data sets. In order to confirm that the records are statistically different, we performed a t-test on each set of data. These results showed convincingly that the hypothesis fails the t-test with t=-2.15 (p=0.038, Tc=2.02). We note that removing the three discrepant results would give a general record in better agreement with our results, although this assumes that the wingnut specimen is correctly placed in age. Because the Chinese wingnut specimen was dated with $^{14}$C wiggle-matching, our results also covered two decades prior to 3371 BCE which might include possible shifts in the calendar dates which could be derived from the simple wiggle-match. Nevertheless, we do not observe any excursion which might be consistent with Wang *et al.* [10].

Discussion

The results presented here raise some important points regarding radiocarbon and dendrochronological dating. Wang *et al*. [10] stated they applied the standard method of dendrochronological dating to their specimen. They claim to use crossdating and correlated the wingnut rings with a "master chronology of tree-ring widths from California" downloaded from the International Tree-Ring Data Bank (ITRDB at ncds.noaa.gov).We assume they used the record of bristlecone pine from the White Mountains that spans 7,091 years (5142 BCE - 1962 CE) (https://www.ncdc.noaa.gov/paleo-search/study/3254). However, the dating approach applied by Wang *et al*. [10] is far from conventional dendrochronology. The annual ring growth variations of bristlecone pine from the alpine environment of the White Mountains are limited by the cold and extremely dry climate [19]. They are not comparable to Chinese wingnut (*Pterocarya stenoptera*) which has a completely different ecological amplitude, different climatic growth controls, and grows on the other site of the world. Additionally, the wingnut sample contains only ~60 tree rings, is from a tree species which may contain interannual density fluctuations and/or missing rings, and is only to a very limited extent suitable for dendrochronological dating purposes. There is no reasonable ecological reason for California bristlecone pine ring widths to be significantly correlated (0.525) with ring widths from a Chinese deciduous tree from humid subtropical Hubei province in Southeast Asia. In dendrochronology, crossdating on radial ring growth is performed only between trees whose growth is limited by the same environmental stimuli, in most cases regional climatic conditions. Without the existence of an, as yet unknown, climate teleconnection, the patterns of wide and narrow radial-growth (rings) formed in different climatic conditions (in this case humid-warm versus cold-dry) would be unmatchable. Moreover, there should be no significant correlation between a 2-year lagged time series. Further, Wang *et al*. [10] used the computer program dpLR "to perform the dendrochronology". The dpLR (Dendrochronology Program Library) is a software package and typically is not used to measure ring parameters, nor for correlation analyses of different tree-ring records. More detail in the Wang *et al*. [10] dendrochronological methods is required. Were the ring widths measured? Why did they find it appropriate to crossdate them against California bristlecone pine? We believe it is important for tree-ring crossdating to be carried out according to known and well established methods and practices [20].

    The wingnut specimen was additionally dated with $^{14}$C prior to conducting the high-resolution $^{14}$C measurements for the spike study. According to [10], the wingnut tree rings were counted and four 5-year ring groups sampled for wiggle matching. Wiggle-matching is a technique to match knowingly spaced-age differences (e.g. ring groups) via Monte Carlo simulation of Chi-squared fit of the $^{14}$C data to the IntCal curve [13]. The 4-point sequence of 5-year groups and mismatch to IntCal mean that the wiggle-match of [10] is not reliable. Based on our analysis of the record of [10], although it appears that although the dating of is probably reasonably accurate based on their $^{14}$C wiggle-match analysis, it is likely not accurately dated to the calendar year. The $^{14}$C variation observed by [10] could deviate from their assigned age by several decades or more. Our analysis of a fit to the results of [10] using the D-Sequence function of OxCal [21] places the older end of the wingnut sequence to 3496-3458BC rather than the 3388BC. We do observe an apparent 11-yr Schwabe cycle in our data, as shown in a wavelet analysis in figure 2. This cycle shows a variation in $^{14}$C of up to ~5 per mil over one solar cycle. This is somewhat larger than that observed in more recent trees [9], although similar to the solar-cycle effects observed by Jull *et al*. [15]. Further, as we have already noted, the "event" observed in the Wang *et al*. [10] data only has three years (3370, 3371, 3388 BCE) which are actually discrepant from our data by >2σ arguing against a statistically-significant excursion at that time.

Conclusion.

We do not confirm that the $^{14}$C event at 3372-3371 BCE described by Wang *et al*. [10] occurs in independently-derived tree-ring records. Therefore, unless this "event" can be independently confirmed in other trees or other proxy records for cosmogenic isotopes [8, 9], this result must be excluded from any list of Solar Proton Events (SPE) events. We must also conclude that this "event" does not occur at 3372-3371 BCE. The result presented by Wang *et al*. [10] may either be at a different time-period or may be due to unaccounted measurement effects that have not been fully evaluated. In particular, the effects of the storage of the wood in a reducing river sediment environment may be important. Since our results are generally consistent with the trend of Wang *et al*.'s [10] data with the exception of the two years (3370 and 3371 BCE) and one other year (3388 BCE), we may also conclude that the result are actually in general agreement, but that errors in Wang *et al*. [10], who used two different laboratories, have been underestimated. In this interpretation, the two sets of results are consistent and show the modulation of the $^{14}$C signal by the solar cycle. Only 3371 BCE diverges significantly, which in any case would not be sufficient to claim an SPE event. This underlines the necessity of an abundance of caution in interpreting isolated deviations from a general trend as a significant event. Results need replication in independent records and at different laboratories to ensure reproducibility of the data.

Methods.

The bristlecone wood specimen used in this study (Fig. 3) was collected by M. Salzer in July 2003 at the elevation of 3575 m asl on the Sheep Mountain ridge (37.53475 N and 118.20045 W). The ring widths of this specimen were measured on a Lintab system. The 328-year ring series is correlated with the 6422-yr master chronology from Sheep Mountain spanning from 4408 BCE to 2014 CE, which is composed of tree rings from remnant wood, and overlaps with rings from living trees [16]. The Sheep chronology also correlates well with two other long bristlecone tree-ring chronologies from the same mountain range confirming the dating accuracy the chronology. Dating SH146-2003 was assigned using correlation of ring widths between 50-year segments and the master SHP mountain chronology lagged successively by 25 years with statistical verification using COFECHA12K version 6.06P which uses multiple parametric statistics to examine the quality of various fits [20]. Second checks on dating were confirmed through the presence/absence of subannual ring features such as frost rings. Further details of master chronology construction are published in [16,19]. The crossdating and the master chronology development were performed at the Laboratory of Tree-Ring Research, University of Arizona. The SH146-2003 remnant specimen has no locally absent rings and includes a very prominent frost ring formed in 3458 BCE that is also present in the same year in one other sample of the master chronology. The annual growth layers were separated through cutting with a razor blade, and were grounded to 20 μm-mesh. Each powdered sample was converted to α-cellulose using standard procedures [22, 23]. Bristlecone cellulose samples were combusted to $CO_2$ and converted to graphite and $^{14}$C dating was performed using the 200kV MICADAS at the Institute of Nuclear Research in Debrecen, Hungary [22]. Sample calculation and data reduction were done using the standard BATS software [24].

The oak sample (CHEY1-17) originated from accumulation of 17 subfossil oak trunks found during gravel extraction in the Moselle River valley (48.9405N, 6.0586E) in 2011 (Fig. 2, Supplementary. Tree-ring widths were measured to an accuracy of 1/100 mm using a stereo microscope and a measuring system. The annually resolved tree-ring width series of 14 tree trunks are synchronized into a 247-year mean chronology. This site chronology is overlapped with the master oak chronology for South Germany [16]. After 10-year smoothing spline transformation these two chronologies correlate at r=0.53 for the common period from 3405 BCE to 3159 BCE. The crossdating was performed at University of Freiburg. Samples were converted to α-cellulose using the procedures described in [22] and the AMS samples were converted to graphite using the AGE-3 system at ETH Zürich. Samples were measured on a MICADAS 200kV AMS of the same design and software [23] as the Debrecen machine.

Acknowledgements

This work was supported by the European Union and the State of Hungary, co-financed by the European Regional Development Fund in the project GINOP-2.3.2.-15-2016-00009 'ICER' and the US National Science Foundation P2C2 program (EAR1203749). WT was supported by the German Research Foundation (DFG, TE 613/3-2).

Author contributions

AJTJ, IP, MWS and WT wrote the paper and interpreted the results. NB, MM, TV and LW made the measurements and derived the results. TV, WT and IP provided the figures. IP, CB, MWS and WT provided the absolutely dated tree rings via methods of dendrochronology.

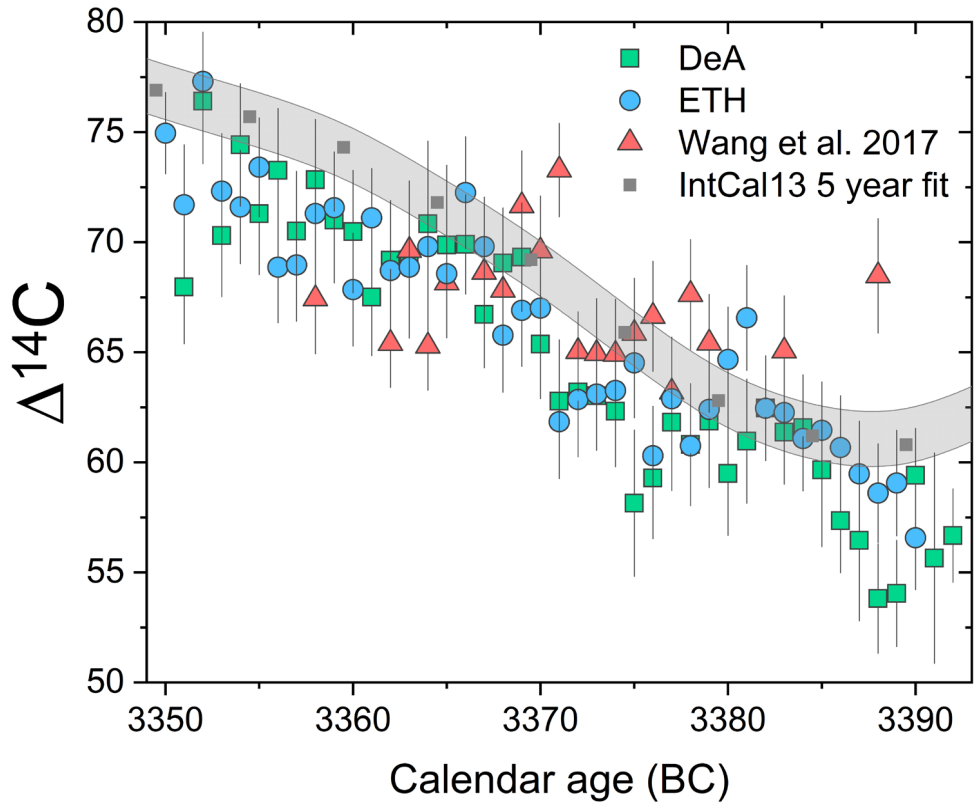

Figure 1: Measurements of $\Delta^{14}C$ (in ‰) vs. dendrochronologically –derived calendar age for samples of Bristlecone Pine (measured in Debrecen, DeA) and European Oak (measured in Zürich, ETH), compared to the reported $\Delta^{14}C$ results for Chinese wingnut of Wang *et al*. [10]. In the case of the wingnut samples, the age reported by [10] is used as the age on the ordinate axis. The gray area shows the smoothed fit of the international calibration curve (IntCal13) [12] with 1σ error.

Figure 2: Wavelet analysis of the Bristlecone Pine $^{14}$C series. Continuous wavelet transform (CWT) shows various variance components of the annual series scaled with a color code (scale bar on the left). The high frequency attributed to solar variability is marked with red fields. Black line outlines the variance significant relative to red noise. A similar result can be obtained for the European Oak series (Figure 1 in Supplementary).

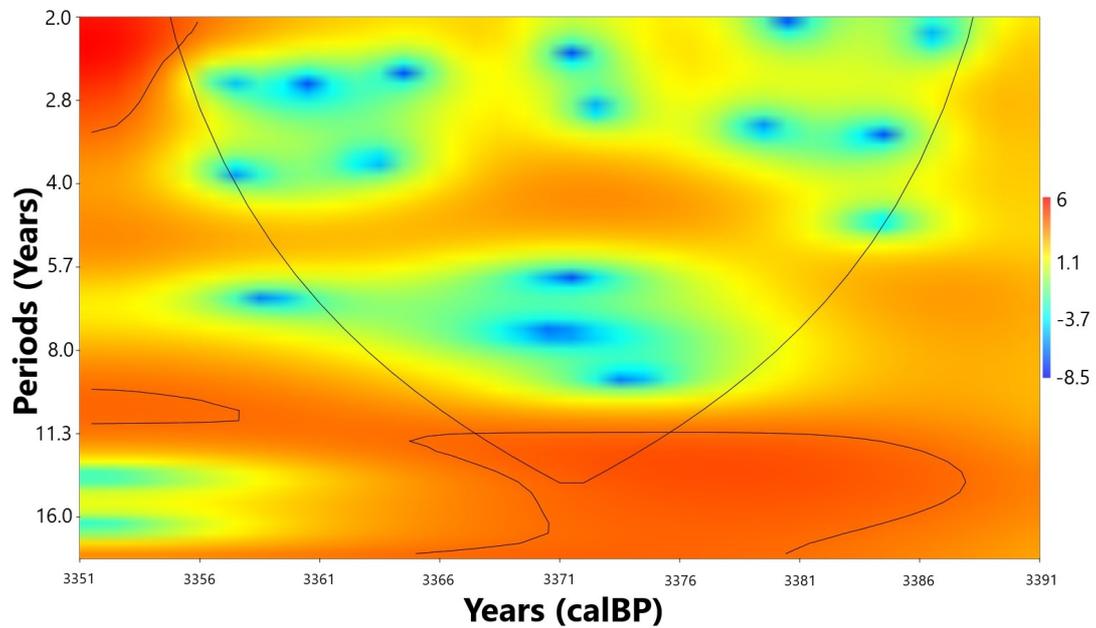

Figure 3: Bristlecone pine (*Pinus longaeva*) specimen SH146-2003 (top) used in this study for cross-dating of the testing tree rings and Chinese wingnut (*Pterocarya stenoptera*) specimen used by Wang et al. 2017 (bottom).

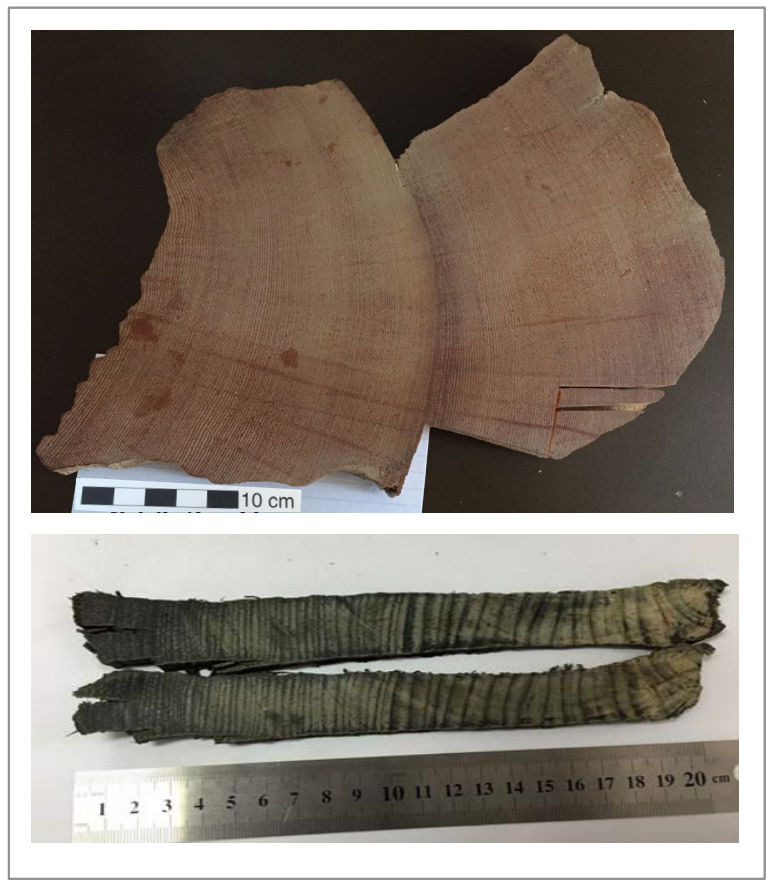